# GYROMONOTRON AS AN INSTRUMENT FOR ELECTRON BEAM COOLING IN ACCELERATORS

V. Khoruzhiy, Kharkov Institute of Physics and Technology, Kharkov, Ukraine

*Abstract*

A gyromonotron is based on energy transformation of the transverse motion of beam particles into electromagnetic wave radiation. This property is proposed to be used for electron beam cooling in the accelerator. The energy of the transverse beam motion is converted into stimulated oscillations of the $H_{111}$ fundamental mode in the gyromonotron resonator. The $H_{111}$ mode excitation is the result of simultaneous excitation of $H_{111}$ and $E_{010}$ modes in the resonator of the given length L, where L is determined by by the start-oscillation condition for certain values of transit angle $\theta_0$ in a monotron. The $H_{111}$ mode actually determines the amplitude of the pump mode, which exists at whatever value of transverse velocities of the beam. We have determined basic gyromonotron parameters such as the radius R and length L of the resonator, and the amplitude of the guiding magnetic field $H_z$ for intended radio frequency $f_1$.

## INTRODUCTION

We propose here cooling of electron beams in in synchrotrons [1-3] (storage rings [4, 5]) using the gyromonotron [6-9] as a device for converting transverse beam energy into the oscillation energy of electromagnetic wave. The gyromonotron is the simplest gyro-oscillator among other similar devices (gyrotron, gyromonotron, gyro-BWO) used for this purpose. The gyromonotron is a empty cylindrical resonator placed in a longitudinal magnetic field of the solenoid.

We suggest effective cooling of electron beam for energy $W_0$ of electron beam more than some hundreds MeV, when starts action quantum fluctuations of macroscopic electron trajectories in accelerators (Ternov-Sokolov effect [10])

$$W_0 > mc^2 (2\pi mcR_s / h)^{1/5}$$

where $h/2\pi$ is reduced Planck constant, $R_s$ is synchrotron's radius.

Expression for $W_0$ can be rewritten in a more convenient form for electrons

$$W_0(MeV) > 151.82(R_s)^{1/5}$$

where synchrotron's radius dimension is $[R_s]$=meter.

Quantum fluctuations (recoil of electrons) lead to growth of emittance through additional particles divergence, hence, as result, growth beam dimensions.

## GYROMONOTRON PARAMETERS

The energy of the transverse motion of the electron beam is converted into the electromagnetic wave energy during multiple passages of the same bunches through the gyromonotron at the given radio frequency $f_1$. The output window for RF power is absent in our case. Let the frequency $f_1$ be the minimal frequency ($H_{111}$ mode or $E_{010}$ mode) excited in the gyromonotron resonator. Let us consider possible gyromonotron parameters.

As is well known [11], at the given frequency $f_1$ (and the corresponding wavelength $\lambda_1$) of $H_{111}$ mode it is possible to determine the radius of the resonator as

$$R_1 = 1.841 / \sqrt{(2*\pi/\lambda_1)^2 - (\pi/L)^2} ; \quad (1)$$

The $H_{111}$ mode is fundamental, if the resonator length is

$$L/(2*R_1) > 1.015$$

The resonator radius for the $E_{010}$ mode is given by

$$R_0 = 2.405 * \lambda_0 / (2*\pi) \quad (2)$$

Radius $R_0$ (2) is independent from the resonator length L. $E_{010}$ is the fundamental mode at the resonator length

$$L/(2*R) < 1.015$$

The necessary condition for gyromonotron operation is the presence of not only transverse velocities of electrons, but the transverse electric field, too. In our case, among the lowest-frequency modes, it is the $H_{111}$ mode that has the necessary electric field $E\varphi$. However, the generation of high-frequency oscillations by the gyromonotron requires the initial transverse velocities to be of the same order of magnitude as the longitudinal velocities are. Then, we have to provide the monotron [12] condition for simultaneous excitation of $H_{111}$ and $E_{010}$ modes as a result of radiative instability [13,14].

This, in turn, requires realization of the condition for excitation of the mode $E_{010}$ with frequency $\omega_0$ in the monotron resonator of length L.

The transition angle for the mode $E_{010}$ in the monotron is [12-14]

$$\theta_0 = (2n+0.5)\pi = \omega_0 L/V_z, \; n=1, 2$$
$$\omega_0 = 2.405*c/R_0$$

(n is the number of the generation zone, $V_z$ is the longitudinal velocity of the beam, c is velocity light). Then the resonator length is

$$L=(2n+0.5)\lambda_0 V_z/2c \quad (3)$$

If the frequency difference $\omega_1 - \omega_0$ for modes $H_{111}$ and $E_{010}$ correspondingly is less than a linear increment of radiative instability, then simultaneous generation of modes $H_{111}$ and $E_{010}$ is possible. Since this instability refers to the Raman-type instability, the increment may be less than the plasma frequency or, at best, equal to the plasma frequency of electron beam $\omega_b$ [14]; then

$$\omega_b > \omega_1 - \omega_0 \quad (4)$$

Expression (4) may be rewritten as

$$n_b > ((\omega_1 - \omega_0)/(5.64*10^4))^2 \quad (5)$$

where $n_b$ is a beam density

Another condition for excitation of the two modes with the same increment is [13]

$$\Delta\theta = \theta_0 - \theta_1 < \pi,$$

where $\theta_1$ is the transit angle for the $H_{111}$ mode with frequency $\omega_1$. As a result, the generation of $E_{010}$ mode is possible due to the conversion of the longitudinal motion energy into the energy of HF oscillations in the monotron. Under realization of the above conditions, the $H_{111}$ mode can be simultaneously excited as a result of mode competition. In our case, the $H_{111}$ mode is in fact a pump mode leading to electron beam cooling under coherent beam radiation. In case of necessity, we can change the amplitude of the $E_{010}$ mode through some change in the resonator length within the zone of generation n.

With the given wavelength $\lambda_1$, it is possible to determine the longitudinal magnetic field strength of the solenoid

$$H_{Z0}(kOe) \simeq 10.7/\lambda_1$$

for a nonrelativistic beam (dimension $[\lambda_1]$ = cm).
For gyromonotron operation at a given frequency $f_1$, the constancy of relativistic cyclotron frequency is required

$$m*\Omega_{c.rel}/(2*\pi) \approx f_1; \quad m = 1, 2...$$

At the same time, the phase focusing calls for the fulfillment of the condition [1]

$$f_1 - m*\Omega_{c.rel}/(2*\pi) > 0$$

The m>1 value (m is the cyclotron frequency harmonic) is used for a high-energy relativistic beam.

As an example, we put the bunch repetition frequency to be $f_1$=350MHz of $H_{111}$ mode (former LEP collider), the corresponding wavelength is

$$\lambda_1 = 85.7 cm$$

then for the generation zone with n=1, using the method of successive approximations, it is possible to determine the resonator radius from (1). Then from formula (2) we determine the wavelength $\lambda_0$, and after that, with the use of eq. (3), we can obtain the resonator length L. By substituting the calculated L value into (1) we eventually find the resonator radius to sufficient accuracy. For the $H_{111}$ mode the gyromonotron radius is found to be R=28.3cm. The resonator length is L=92.5cm (L/2R)>1). This length corresponds to the middle of the first generation zone (n=1, $\theta_0$=2.5$\pi$) of the monotron. The beam density from (5) is $n_b > 3.78*10^7$ cm$^{-3}$. The oscillation frequency of $E_{010}$ mode is $f_0$=405.2MGz, relative detuning is $(f_0-f_1)/f_0$=13.6%. The difference between the transit angles of competing modes is $\Delta\theta = \theta_0 - \theta_1$=1.1<$\pi$, and that provides simultaneous excitation of $H_{111}$ and $E_{010}$ modes. For nonrelativistic beam energies the longitudinal magnetic field strength of the solenoid is

$$H_{Z0} = 0.125 kOe$$

for nonrelativistic beam energy. Conservation of relativistic cyclotron frequency is realized under the condition

$$H_{Z,rel} = \gamma H_{z,0}/m$$

where $\gamma$ is the relativistic factor.

When $\gamma \gg 1$, value of the longitudinal magnetic field $H_{Z,rel}$ can be reduced by using the cyclotron harmonic m>1.

## CONCLUSION

A possibility of using gyromonotron for cooling the electron beam in cyclic accelerators has been considered here. The gyromonotron length is determined by the condition of $E_{010}$ mode excitation through loosing the longitudinal beam energy in the monotron resonator. It has appeared possible to create the condition of mode competition for simultaneous excitation of $E_{010}$ mode and an additional $H_{111}$ mode. $H_{111}$ mode makes possible stimulated cooling of the electron beam with initial transverse velocities $\beta_\perp < 1/\gamma$. The $H_{111}$ mode actually determines the amplitude of the pump mode, which existis at whatever value of transverse velocities of the beam.

Preliminary investigations of the above described device for beam cooling may be carried out using the beam from a electron linac (or an electron gun) as it passes through the giromonotron. Computer simulation of beam cooling is needed as part of preliminary investigations.